\documentclass[final,5p,times,twocolumn]{elsarticle}
\usepackage{graphicx}
\usepackage{bibentry}
\usepackage{subfigure}
\usepackage{epstopdf}
\usepackage{color}
\usepackage{multirow}
\usepackage{gensymb}
\usepackage{lineno}
\usepackage{epsfig}
\usepackage{amsmath}
\usepackage{amssymb}

\biboptions{square,comma,sort&compress}
\journal{Modelling and Simulation in Materials Science and Engineering}

\usepackage[colorlinks=true,citecolor=blue,linkcolor=blue,urlcolor=blue]{hyperref}

\usepackage[table,xcdraw]{xcolor}
\usepackage{graphicx}
\usepackage{listings}
\usepackage{booktabs}
\renewcommand{\BibitemShut}[1]{}
\newcommand{\mybox}[1]{\colorbox{black}{\textbf{\textcolor{white}{#1}}}}
\newcommand{\mygbox}[1]{\colorbox{lightgray}{\textcolor{black}{#1}}}
\begin{document}

\title{$\mu$2mech: a Software Package Combining Microstructure Modeling and Mechanical Property Prediction}

\author[a]{Albert Linda}
\author[a]{Ankit Singh Negi}
\author[a]{Vishal Panwar}
\author[a]{Rupesh Chafle}
\author[a]{Somnath Bhowmick}
\author[b]{Kaushik Das}
\author[a]{Rajdip Mukherjee\corref{cor}}
\ead{rajdipm@iitk.ac.in}

\address[a]{Department of Materials Science and Engineering, Indian Institute of Technology Kanpur, Kanpur 208016, UP, India}
\address[b]{Department of Metallurgy and Materials Engineering, Indian Institute of Engineering Science and Technology, West Bengal, Shibpur, Howrah, 711103, India}

\cortext[cor]{Corresponding author(s). Tel.:+91 512 259 6449; Fax:+91 2597505 } 

\begin{abstract}
We have developed a graphical user interface (GUI) based package $\mu$2mech to perform phase-field simulation for predicting microstructure evolution. The package can take inputs from \textit{ab initio} calculations and CALPHAD (Calculation of Phase Diagrams) tools for quantitative microstructure prediction. The package also provides a seamless connection to transfer output from the mesoscale phase field method to the microscale finite element analysis for mechanical property prediction. Such a multiscale simulation package can facilitate microstructure-property correlation, one of the cornerstones in accelerated materials development within the integrated computational materials engineering (ICME) framework. 
\end{abstract}
\maketitle
\section{Introduction}
The age-old practice of trial and error-based experimental methods for developing new materials is being replaced by ICME-based methods. The latter reduces the cost and time required for developing new materials. While designing a new alloy, predicting its microstructure is one of the most crucial steps. Although elastic stiffness constants of materials are intrinsic (as they depend on bonding), the microstructure can be controlled by processing conditions, allowing the tuning of effective mechanical properties. Thus, a package combining microstructure modeling and mechanical property prediction can be convenient for alloy design.

Phase field simulation is the most popular technique used for microstructure modeling.\cite{book_pf,pf_review} In recent years, phase field simulations have been used to simulate complex microstructures evolution during different processes, such as solidification~\cite{chatterjee2008phase, HOTZER2015194, ZHAO20191044,FU2017187}, spinodal decomposition~\cite{bhattacharyya2003study,ramanarayan2003spinodal,CHAFLE1}, precipitate/grain growth~\cite{Bandyopadhyay2023,mukherjee2009, Mukhrjee2010, CHANG201767, PhysRevLett.86.842, verma2021grain, SITOMPUL2022101834}, coarsening~\cite{Mukherjee2013,MOLNAR20126961}, and the effect of an external field\cite{Gururajan2007, CHAFLE}.

One popular tool researchers utilize is MOOSE (Multiphysics Object Oriented Simulation Environment) \cite{lindsay2022moose}. This powerful software offers a range of capabilities for conducting complex multiphysics simulations. MOOSE is a flexible and extensible platform that empowers the development of phase field simulations. Its application extends to the study of microstructural evolution and dendritic growth\cite{JOKISAARI2018336}. Another valuable resource in this field is OpenPhase \cite{TEGELER2017173}, a library dedicated to implementing the multiphase field method for simulating microstructure evolution during materials processing. OpenPhase excels in various applications, including grain growth and solidification, while offering high-performance capabilities through distributed memory simulation. The "Parallel Algorithms for Crystal Evolution in 3D" (PACE3D)\cite{HOTZER20181}, offers an integrated solution for addressing multi-physics applications in a parallel and efficient manner. The solver architecture within PACE3D encompasses diffuse interface methods, grain growth, grain coarsening, solidification, fluid dynamics, mechanical interactions, and  electrochemistry. Another recently developed tool is MicroSim\cite{microsimMicroSimMicrostructure}; it comprises multiple modules, encompassing a grand-potential based solver, Kim Kim Suzuki (KKS)\cite{PhysRevE.60.7186} models for OpenCL\cite{5457293} and CUDA\cite{7476520}, a Cahn-Hilliard model with FFTW, and OpenFoam-based solvers\cite{openfoamOpenFOAMOpenFOAM} for multiphysics simulations. MicroSim incorporates a user-friendly graphical interface that enables users to easily create input files, select solvers, and visualize simulation results using Paraview\cite{paraviewParaViewOpensource}. Despite the advantages provided by these tools, certain limitations hinder their accessibility. MOOSE and OpenPhase, for instance, have steep learning curves, which can pose challenges for programmers or researchers seeking user-friendly interfaces for their simulations. Additionally, the absence of built-in graphical user interfaces (GUIs) to visualize output in these tools can be cumbersome for new users.

In studying microstructures and their impact on material properties, OOF2 (Object-Oriented Finite Element Analysis)\cite{OOF2} stands out as specialized software. By employing finite element analysis techniques, OOF2 enables researchers to simulate the intricate behavior of complex structures at the microscopic level. Its user-friendly interface facilitates the definition of material properties, boundary conditions, and loading conditions. Moreover, OOF2 offers a comprehensive range of features for analyzing and visualizing microstructures, encompassing grain boundaries, phase distributions, and stress/strain fields. By simulating the relationship between microstructure and mechanical properties, OOF2 facilitates investigations into the influence of diverse microstructural features on material behavior. Two other prominent software packages commonly employed for finite element analysis and microstructural studies are ABAQUS\cite{abaqus} and ANSYS\cite{ansys}. ABAQUS, a widely used commercial software, offers advanced capabilities for investigating the mechanical properties of materials, including microstructures.
On the other hand, ANSYS encompasses a suite of simulation tools, including finite element analysis, with specific modules such as ANSYS Mechanical and ANSYS Multiphysics catering to the study of microstructural behavior. These software packages enable researchers to simulate material deformation, phase transformations, and grain boundary effects. Notably, both ANSYS and ABAQUS provide the ability to perform finite element analysis of microstructures, incorporating crystal plasticity through the use of CPFEM (Crystal Plasticity Finite Element Method)\cite{cpfem_book} by utilizing their respective subroutines.

Microstructure-based mechanical property prediction involves scale bridging, from mesoscale phase-field modeling to macroscale finite element analysis. While stand-alone packages are available for both the length scales, a package combining them does not exist, to the best of our knowledge. Motivated by this, we develop $\mu$2mech software. It can simulate the microstructure of materials using phase-field techniques. The software uses various open-source libraries for numerical computation, visualization, and data analysis. The package can take inputs from \textit{ab initio} calculations, and CALPHAD\cite{saunders1998calphad} software like ThermoCalc \cite{ANDERSSON2002273} for quantitative microstructure prediction. Its features include the ability to simulate binary microstructures for spinodal decomposition and precipitate growth, visualization of microstructures in 2D and 3D, and calculating particle size and volume. Additionally, $\mu$2mech can be coupled with OOF2 to study the mechanical behavior of the simulated microstructures.

The paper aims to comprehensively describe the $\mu2$mech package, including a brief description of  CALPHAD integration, numerical implementation of the phase field technique, and installation. We discuss an example of FeCr binary alloy, showcasing the capabilities of $\mu2$mech to simulate the microstructure and its integration with OOF2 for mechanical property prediction. The example of FeCr binary alloy also serves as a manual for users, providing guidelines regarding various inputs to the code and outputs generated. Finally, we discuss the impact that $\mu$2mech can have on the materials science and engineering community.
 
\begin{figure*}[htpb]
    \centering
    \includegraphics[width=0.8\linewidth]{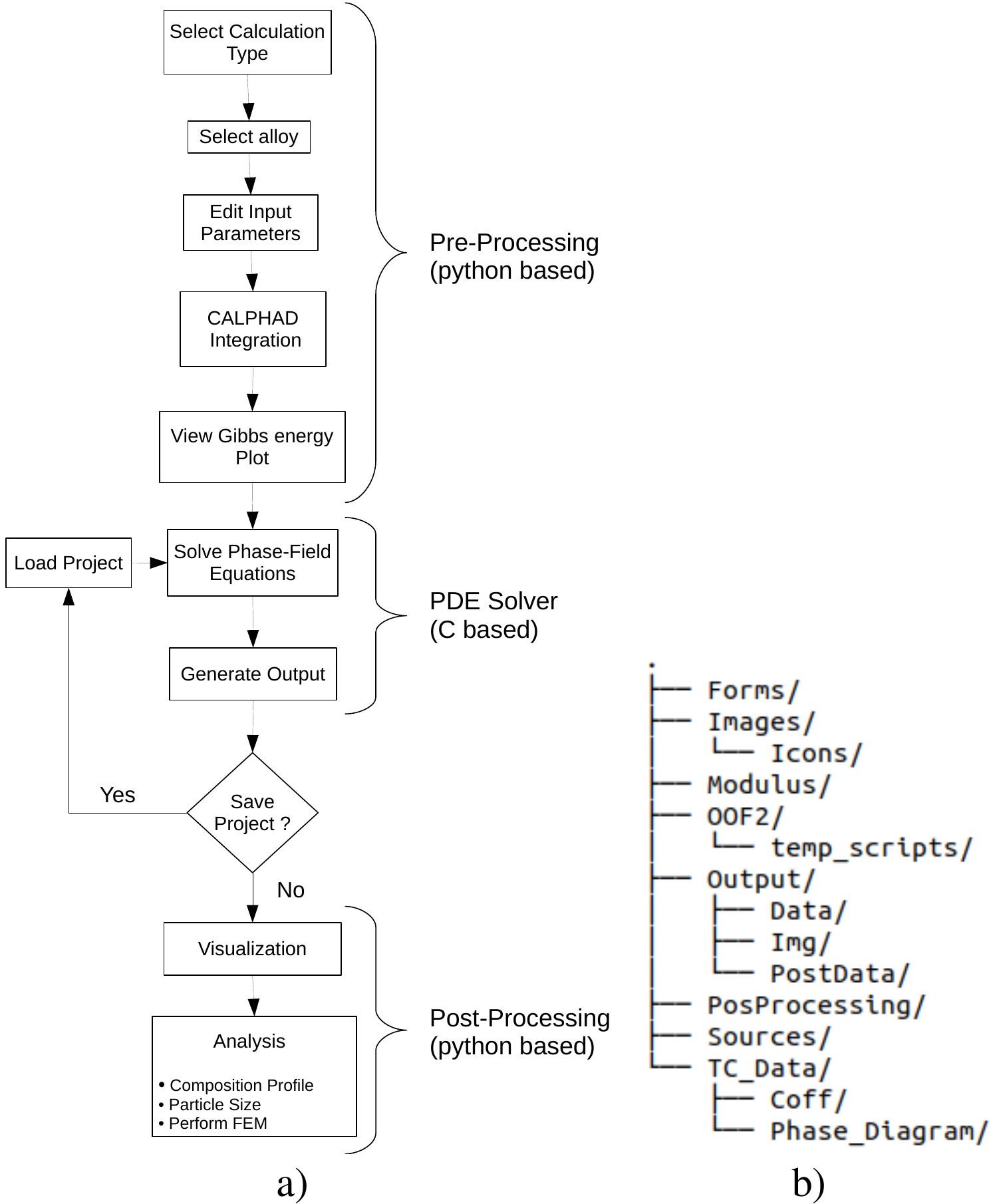}
    \caption{a) Flowchart depicting the complete process flow of $\mu$2mech, encompassing all essential steps. b) Folder structure overview of $\mu$2mech, outlining the organization of directories.}
    \label{fig1}
\end{figure*}

\section{Software description}
The software package comprises three main steps: preprocessing, partial differential equation (PDE) solver, and post-processing. Figure~\ref{fig1}(a) schematically shows the software's architecture, process flow, and capabilities. During preprocessing, users must select the calculation type (2D or 3D) and other input parameters. Then, the software numerically solves the governing PDEs and generates the microstructure evolution as a function of time. In the post-processing stage, users can visualize the evolution of the microstructure and do many analyses (like particle size distribution and composition profile). Finally, users can create input files for finite element analysis via OOF2, allowing researchers to simulate materials' mechanical behavior under various loading conditions. Coupling with OOF2 is chosen as it can account for the material property's nonlinearity by generating a mesh according to the morphology of the underlying microstructure. 

The architecture comprises two main components: GUI (pre and post-processing) and the PDE solver. The latter is entirely written in the programming language C. The discretized equations are solved using the Fourier spectral method, and it employs the FFTW library\cite{Frigo} to perform these calculations. In the following section, we describe some technical details related to the phase-field method, like CALPHAD integration and the numerical scheme to solve the Cahn-Hilliard equation. The GUI component, which provides a user-friendly interface for interacting with the software, is written in Python3 by utilizing the PySide2 framework \cite{PySide2}. The $\mu$2mech package also incorporates several other libraries such as NumPy for matrix manipulation, Matplotlib for plotting 2D microstructure intersections, PyVistaQT for visualizing vtk files, PyYAML for project configuration and saving, and FFmpeg \cite{ffmpeg} for generating microstructure evolution animations. For its coupling with OOF2 \cite{919261, OOF2}, $\mu$2mech prepares an input Python script once the microstructure image is passed along with other input parameters.

Detailed installation instructions are given in Appendix~\ref{secinstall}. Post-installation, the directory structure is illustrated in Figure~\ref{fig1}(b). The C source codes and their compiled binaries for the phase field calculations are found in the ``Sources" folder. A \textit{Makefile} has been provided to recompile the source files. Typically, recompilation is not needed unless modifications have been made in one of the source files. The ``Coeff" sub-folder within the ``TC\_Data" directory encompasses thermodynamic details pertinent to the binary alloys, which are crucial for conducting the phase field calculations. These details are stored in the \textit{csv} format, which follows the naming convention ``A-B.csv", where A and B are the constituent elements. An example for the Fe-Cr alloy is illustrated in Table \ref{ref:tc_table} [Appendix~\ref{seccalphad}]. Contained within the ``Phase\_Diagram" sub-folder of the ``TC\_Data" directory are the phase diagrams, serving as a visual reference and not directly involved in the calculation process. The directory labeled as ``Output" temporarily holds the resulting data from the computations of phase field. The ``Post\_Processing" directory temporarily stores the composition plot generated during the post-processing stage. Other folders like ``Forms," ``Images," ``Modulus," and ``OOF2" are mainly responsible for the GUI part of  $\mu2mech$, and it is recommended not to modify these directories.

\begin{figure*}
\centering
\includegraphics[width=0.6\linewidth]{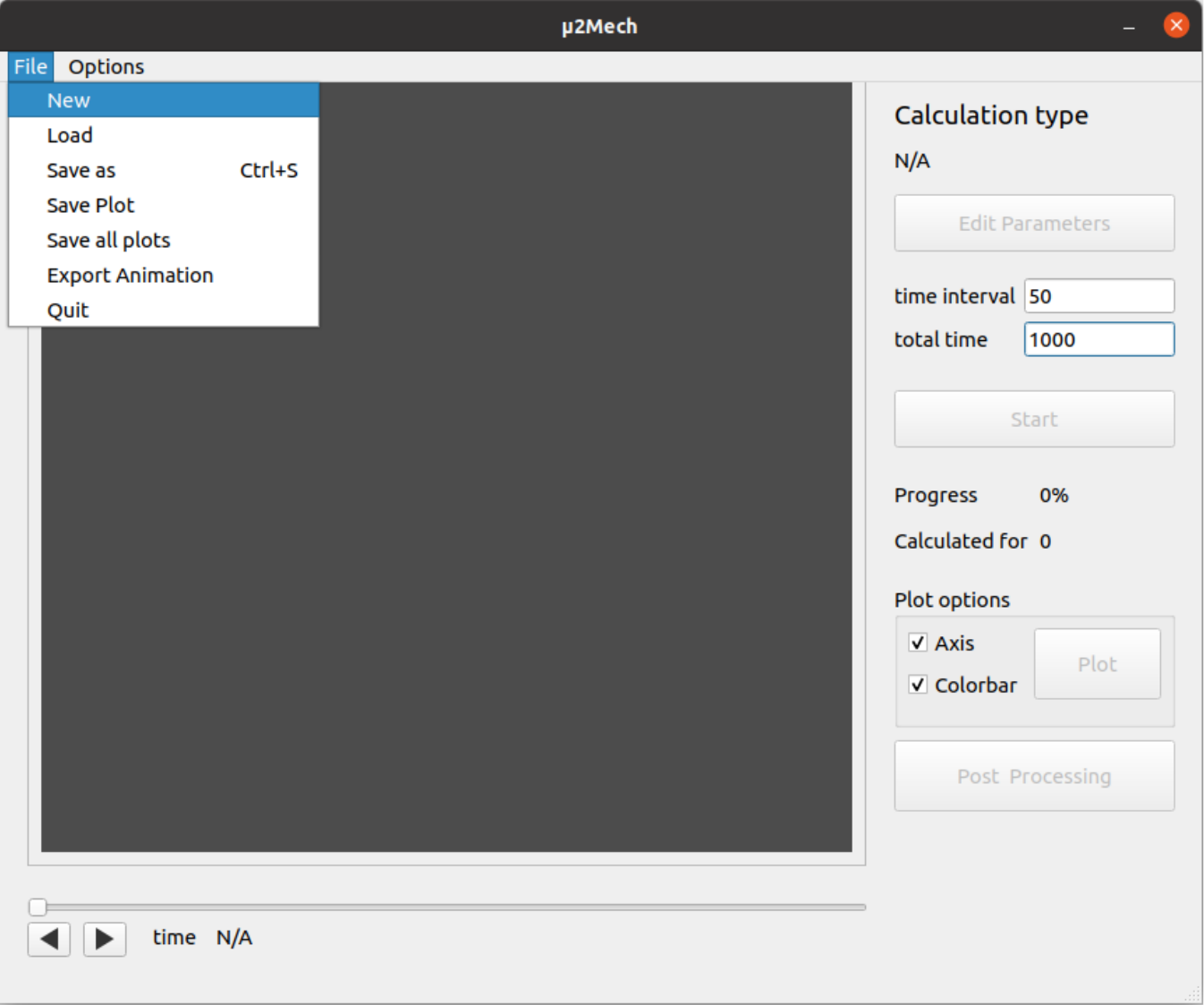}
\caption{Main window of $\mu$2mech.}
\label{ref:local_run}
\end{figure*}


\section{Algorithm: Phase-Field modeling}
\label{secalgo}
The Cahn-Hilliard equation is a fourth-order, nonlinear partial differential equation crucial in studying phase separation and microstructure evolution in multi-component systems. It has been extensively applied to model the time-dependent behavior of various materials, including alloys, polymers, and colloids\cite{Cahn1958}. One can write the Cahn-Hilliard equation as,
\begin{equation}
\label{eq3}
\frac{\partial \phi}{\partial t} = \nabla \cdot \left[ M(\phi) \nabla \left( \frac{\delta F}{\delta \phi} \right) \right],
\end{equation}
where $t$ is the time, $M(\phi)$ is the mobility function, $F$ represents the free energy functional, and $\frac{\delta F}{\delta \phi}$ is the variational derivative of $F$ with respect to $\phi$. The free energy functional typically comprises a bulk free energy term $f_b(\phi)$ and a gradient energy term $\kappa|\nabla \phi|^2$.

The Fourier spectral method is a powerful numerical technique for solving the Cahn-Hilliard equation in 2D and 3D configurations, offering a robust and efficient approach to studying complex microstructure systems\cite{boyd2001chebyshev}. It relies on the Fourier transform to convert the spatial derivatives into algebraic operations in the frequency domain\cite{boyd2001chebyshev}. This approach allows for efficient and accurate computation of the solution. The Fourier spectral method involves discretizing the time variable using a suitable time-stepping scheme, such as the semi-implicit or fully implicit method\cite{EYERT1996271}. Then, the Fourier transform is taken with respect to the spatial variables $x$, $y$, and $z$ (for 3D) or just $x$ and $y$ (for 2D), yielding an equation in the frequency domain,
\begin{equation}
\label{eq4}
\frac{\partial \hat{\phi}}{\partial t} = -M(\phi) \left[ k^4 \hat{\phi} + k^2 \hat{f_b^\prime(\phi)} \right],
\end{equation}
where $\hat{\phi}$ and $\hat{f_b^\prime(\phi)}$ are the Fourier transforms of $\phi$ and $f_b^\prime(\phi) = \frac{\delta f_b(\phi)}{\delta \phi}$, respectively, and $k^2 = k_x^2 + k_y^2 + k_z^2$ (in 3D) or $k^2 = k_x^2 + k_y^2$ (in 2D).

The next step is to solve the frequency-domain equation for $\hat{\phi}(t)$ and update the solution at each time step. Starting from the initial condition $\phi(\mathbf{x}, 0)$ and its discrete Fourier transform $\hat{\phi}(0)$ with respect to the spatial variables, one studies the time evolution by updating $\hat{\phi}(t)$ using an appropriate time-stepping scheme, like the semi-implicit Euler method,
\begin{equation}
\hat{\phi}^{n+1} = \frac{\hat{\phi}^{n} - \Delta t \cdot M(\phi^n) \cdot k^2 \hat{f_b^\prime(\phi^n)}}{1 + \Delta t \cdot \kappa \cdot M(\phi^n) \cdot k^4},
\label{ref:ph_discretized}
\end{equation}
where $\Delta t$ is the time step, and the superscript $n$ denotes the time level and $\kappa$ is gradient energy coefficient . The mobility function $M(\phi)$ can also be incorporated into the time-stepping scheme. Once the updated solution $\hat{\phi}^{n+1}$ is obtained in the frequency domain, the inverse Fourier transform is applied to recover the solution $\phi(\mathbf{x}, t^{n+1})$ in the spatial domain \cite{DFT}. One repeats this procedure until the desired final time is reached.

One can efficiently implement the Fourier spectral method for solving the Cahn-Hilliard equation using the Fast Fourier Transform (FFT) algorithm, which significantly reduces the computational cost compared to traditional finite difference or finite element methods \cite{cooley1965algorithm, Press2007}. Moreover, the spectral method offers high accuracy and preserves the smoothness of the solution, making it well-suited for studying the evolution of complex microstructures \cite{canuto2007spectral}.

\begin{figure*}
    \centering
    \includegraphics[width=0.6\linewidth]{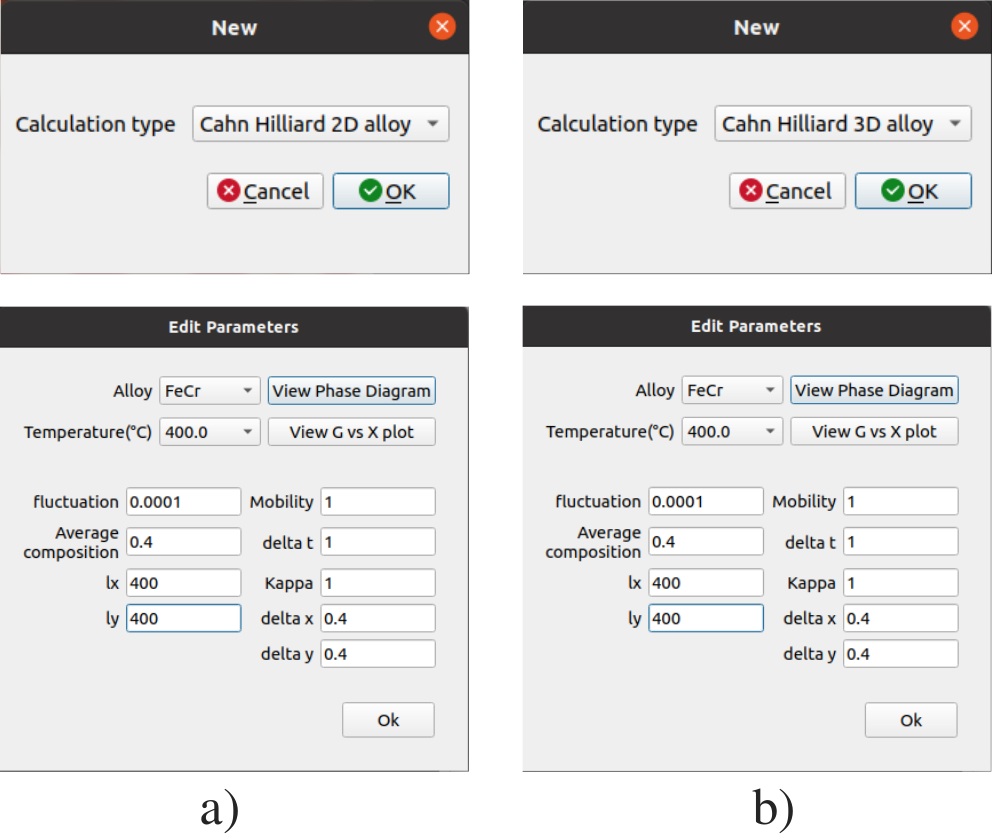}
    \caption{Top panels show the calculation types, (a) 2D and (b) 3D phase field simulations. The bottom panels illustrate the simulation parameters, such as alloy, temperature, average composition, system size, mobility, kappa, and time interval. These parameters play a crucial role in defining the simulation setup and needs to be adjusted carefully.}
    \label{ref:start_calc_edit_parameters}
\end{figure*}

\begin{figure}
    \centering
    \includegraphics[width=\linewidth]{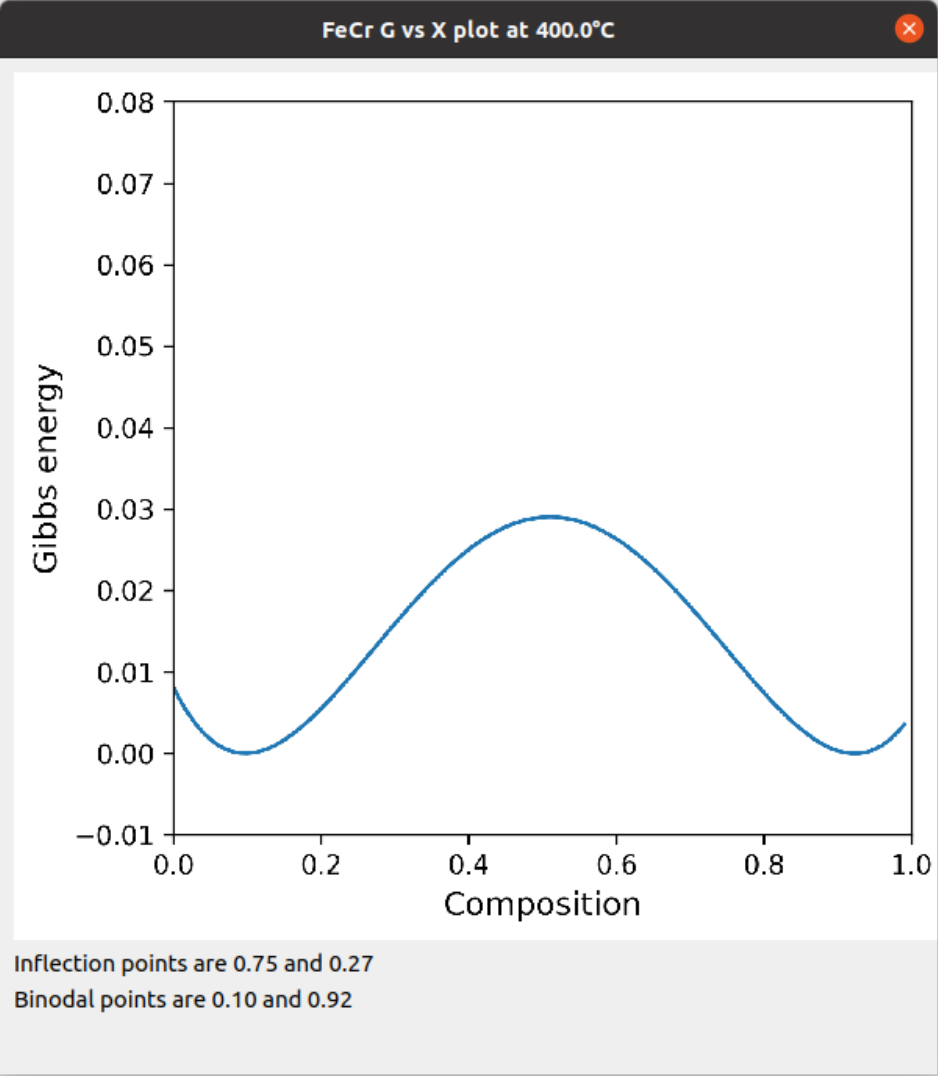}
    \caption{Gibbs Energy versus Composition Plot for Fe-Cr alloy at 400°C. The plot highlights inflection points (0.27 and 0.75) and binodal points (0.10 and 0.92) in mole fraction of Cr.}
    \label{ref:gibbs_plot}
\end{figure}

\section{Illustrative example}
To demonstrate the capabilities of $\mu$2mech, we present an example of a binary system, Fe-Cr, using Gibbs free energy data from CALPHAD calculations [technical details provided in Appendix~\ref{seccalphad}]. The main window of \mybox{$\mu$2mech} is illustrated in Figure~\ref{ref:local_run}. As shown in the diagram, users click the ``New'' submenu under the ``File'' menu to start a new calculation, which opens the \mybox{New} dialog box [Figure~\ref{ref:start_calc_edit_parameters}]. In this dialog box, the users select either \mygbox{Cahn Hilliard 2D alloy} or \mygbox{Cahn Hilliard 3D alloy} from the ``Calculation type$\downarrow$" drop-down menu, as shown in the top row of Figure~\ref{ref:start_calc_edit_parameters}. This action loads the default parameters for calculations. Next, the users can click the \includegraphics[width=0.15\textwidth]{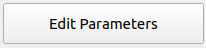} 
button near the top right of the main \mybox{$\mu$2mech} window [Figure~\ref{ref:local_run}]. This action opens the \mybox{Edit Parameters} dialog box, as shown in the bottom row of Figure~\ref{ref:start_calc_edit_parameters}. Next, users select \mygbox{Fe-Cr} from the ``Alloy$\downarrow$'' drop-down menu [Figure~\ref{ref:start_calc_edit_parameters}]. At present, the other available option is \mygbox{Fe-Cu}. Users can add more alloy systems by creating a \textit{newalloy.csv} file in the format specified in Table~\ref{ref:tc_table} [Appendix~\ref{seccalphad}] and saving it in the folder TC\_Data/Coff. To use this new data, users have to restart the application. Users can visualize the phase diagram for the given system by clicking on the adjacent \includegraphics[width=0.15\textwidth]{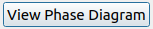} button. The users then select the temperature from the ``Temperature$\downarrow$'' drop-down menu and view the Gibbs free energy versus composition plot by clicking the adjacent \includegraphics[width=0.15\textwidth]{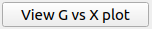} button [Figure \ref{ref:start_calc_edit_parameters}]. A point on the curve fixes the \textit{average composition} for the alloy. When the temperature is 400\degree C for the Fe-Cr system, the spinodal phase forms within the $0.27 \leq X \leq 0.75$ (inflection points on free energy curve, Figure \ref{ref:gibbs_plot}).

If required, the users can modify other parameters before running the simulation [bottom row of Figure~\ref{ref:start_calc_edit_parameters}]. The list of parameters includes ``Fluctuation'' (a Gaussian noise added on top of the average composition to generate the initial composition field), ``lx, ly, lz'' (system size), ``delta x, delta y, delta z'' (grid size), ``delta t'' (time interval), ``Mobility'', and ``Kappa''. The symbols used are consistent with Equations~\ref{eq3},~\ref{eq4},~\ref{ref:ph_discretized}. All the parameters are self-explanatory and consistent with Section~\ref{secalgo}. In this example, we choose lx=ly=400 with delta x=delta y=0.4, which implies a mesh size of $100\times100$. ``Fluctuation'' 0.0001 defines the initial composition field as ``Average composition'' $0.4 \pm $ a Gaussian noise with width 0.0001 at every grid point of the mesh.

After editing the parameters, users return to the main window of \mybox{$\mu$2mech} and set the ``total time'' for the phase-field simulation to run and ``time interval'' after which simulated microstructures are saved for visualization and other analysis. In this example, the total time for microstructure evolution is 1000, and the interval for saving microstructures is 50 [Figure \ref{ref:local_run}]. Next, the users can choose between two options: (i) click the ``Save as'' submenu under the ``File'' menu to save the project, transfer the files and run the calculations on a \textit{remote} machine, such as a high-performance cluster (HPC) or (ii) click the \includegraphics[width=0.15\textwidth]{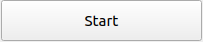} button  to perform the calculations \textit{locally} on a desktop/workstation.  

\begin{figure}
\centering
\includegraphics[width=\linewidth]{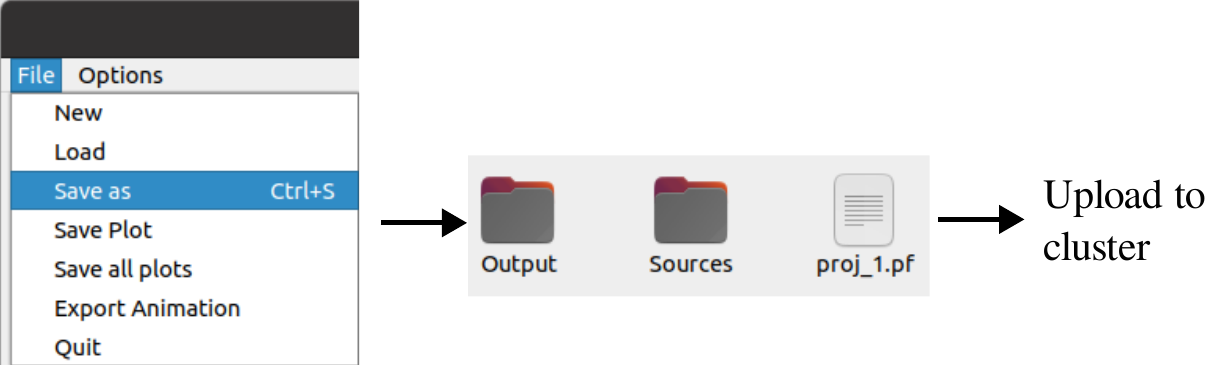}
\caption{The figure illustrate the sequential steps involved in running a calculation in a high performance cluster (HPC).}
\label{ref:server_run}
\end{figure}

To run the calculations in a remote HPC, the user first transfers the entire project folder (saved in the *.pf format) to the HPC [Figure~\ref{ref:server_run}]. Assuming \textit{slurm} workload manager is installed in the HPC, a job submission command may look like \textit{sbatch slurm.sh}, where \textit{slurm.sh} is a bash file containing the scheduler parameters and commands required for calculation. However, the exact command will depend on the installations in an HPC. Once completed, the user transfers the files back to the local machine and uses the ``Load'' submenu under the ``File'' menu in the main \mybox{$\mu$2mech} window to load the finished calculations as a project to do further analysis.

\begin{figure}
\includegraphics[width=\linewidth]{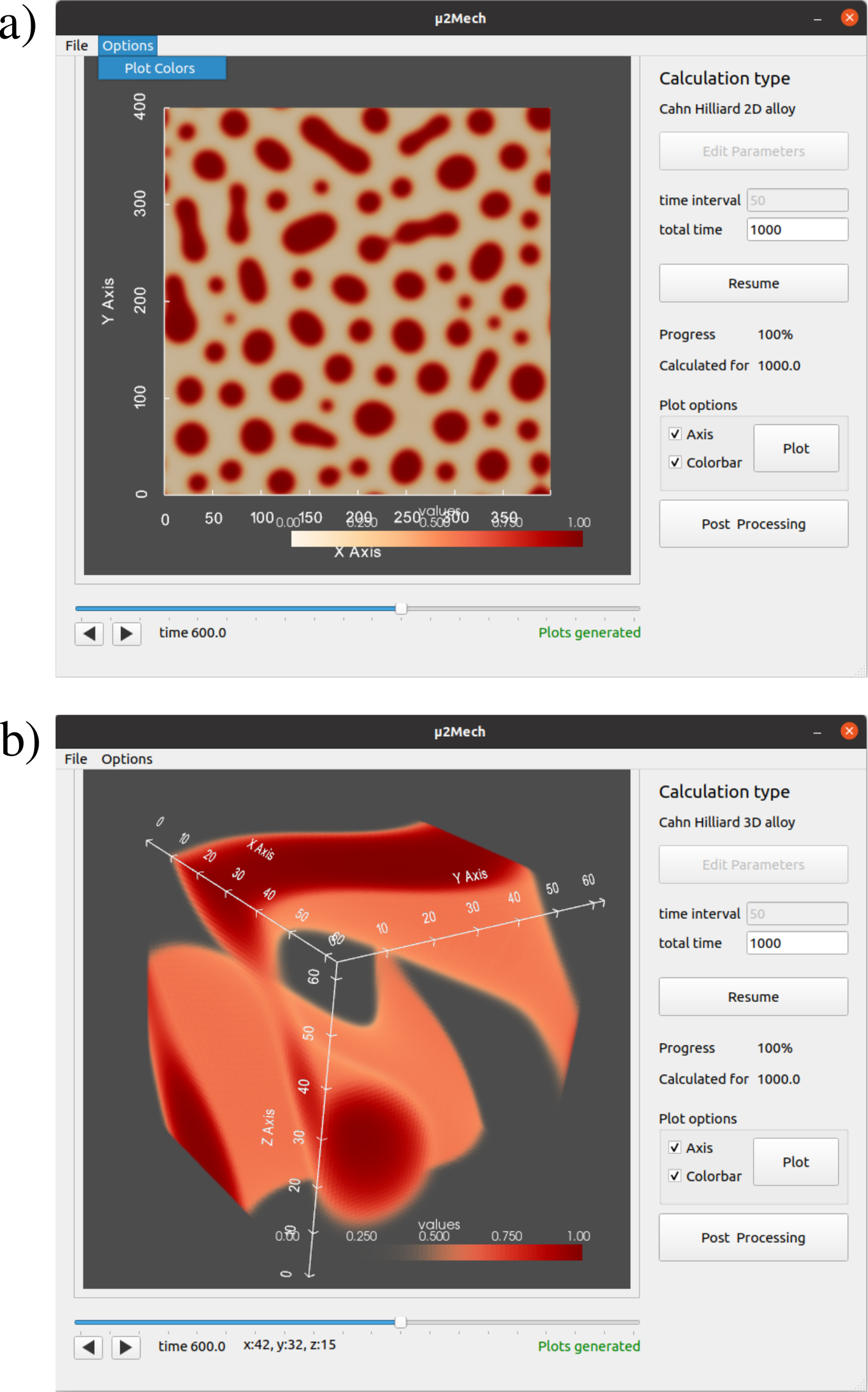}
\caption{a) A 2D and (b) a 3D microstructure of a binary alloy are displayed at the 600th time step, following the completion of calculation. Microstructures at different time steps can be accessed using the time slider at the bottom.}
\label{ref:visualization_2D}
\end{figure}

To run the calculations locally, the user clicks the \includegraphics[width=0.15\textwidth]{Inline_Images/start.png} button in the main \mybox{$\mu$2mech} window [Figure \ref{ref:local_run}]. The status bar below the \includegraphics[width=0.15\textwidth]{Inline_Images/start.png} button indicates the progress of the calculations, and the calculation can be paused and resumed as needed. ``Save as'' and ``Load'' options are available under the ``File'' menu in the main \mybox{$\mu$2mech} window [Figure \ref{ref:local_run}]. After the run, the user can save the project files using the ``Save as'' option, which saves the input and output files. The file extension for saving the project is *.pf. When loading the project, the corresponding *.pf file must be selected. After loading the project, the users can either resume the calculations for more time steps or perform post-processing. To perform additional time evolution, one has to increase the total time and click the \includegraphics[width=0.15\textwidth]{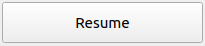} button in the main \mybox{$\mu$2mech} window [Figure~\ref{ref:visualization_2D}].

\begin{figure*}[tphb]
    \centering
    \includegraphics[width=0.75\textwidth]{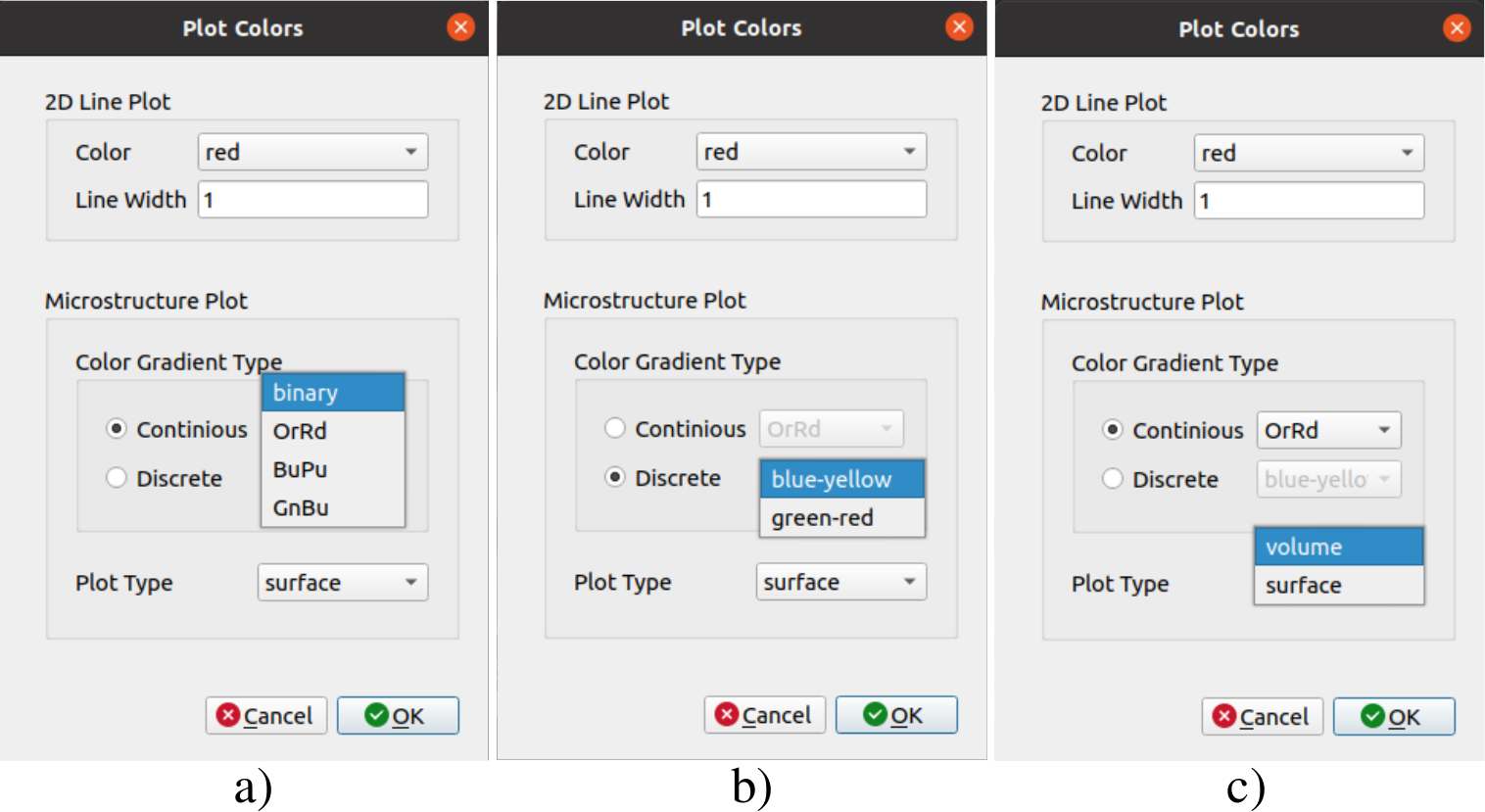}
    \caption{The Plot Colors dialog box for adjusting the appearance of line and microstructure plots. For 2D line plots, color selection is available. In the case of microstructure plots (both 2D and 3D), users can choose between two plot gradient types: continuous and discrete with their color gradient. Additionally, surface/volume plot options are provided for visualizing outer planes and partially transparent isosurfaces, respectively. }
    \label{ref:plot_options}
\end{figure*}

Users can visualize microstructures in the main \mybox{$\mu$2mech} window itself. The microstructures for different time steps are accessible by clicking the left/right arrow key or sliding the slider. Figure \ref{ref:visualization_2D} depict the 2D and 3D microstructure at $600^{th}$ time step. Users can customize the color scheme for the microstructure visualization by selecting the ``Plot Colors'' option within the ``Options'' menu [Figure \ref{ref:visualization_2D}], which opens a new dialog box \mybox{Plot Colors}, as depicted in Figure \ref{ref:plot_options}. The simulated microstructures can be saved (using ``Save Plot'' or ``Save all plots'') or exported as an animation (using ``Export Animation''). These options are available under the ``File'' menu in the main \mybox{$\mu$2mech} window [Figure \ref{ref:local_run}].

\begin{figure}[tphb]
    \centering
    \includegraphics[width=0.5\textwidth]{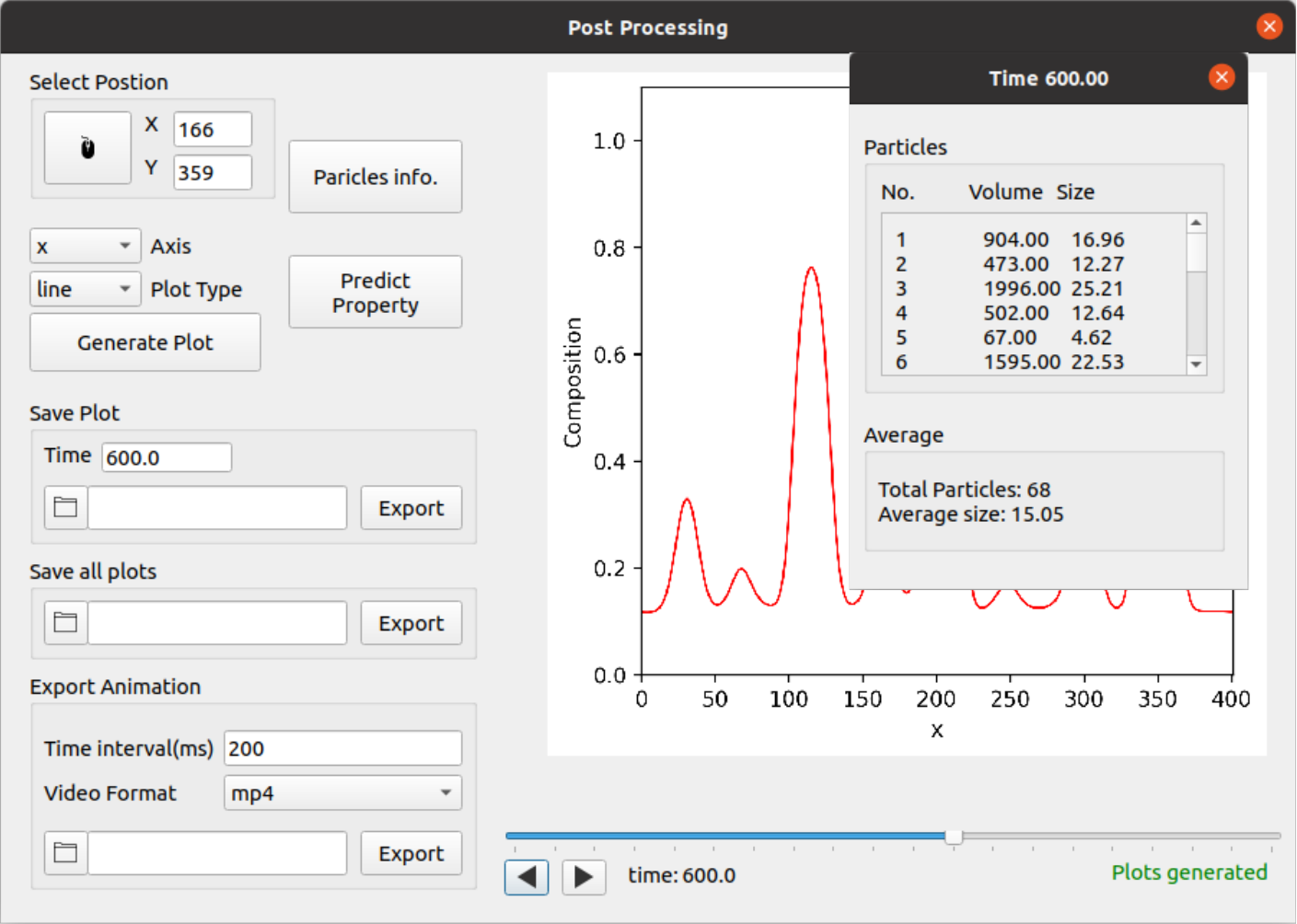}
    \caption{The Post Processing dialog box for 2D microstructure analysis. The main dialog box contains options for composition profile, particle size, and exporting plots and animations related to the composition profile. By clicking the Particles Info. button, a small additional dialog box appears atop the previous one, displaying detailed particle size information.}
    \label{ref:post_processing_1}
\end{figure}

The users can start the post-processing exercise by clicking the \includegraphics[width=0.15\textwidth]{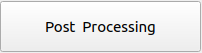} button located at the bottom right of the primary \mybox{$\mu2$mech} window [Figure \ref{ref:visualization_2D}], which opens the \mybox{Post Processing} dialog box, as shown in  Figure~\ref{ref:post_processing_1}. Available options are a) particle size distribution, b) composition profile analysis, and c) property prediction. Particle size distribution at each time step can be analyzed by selecting the \includegraphics[width=0.1\textwidth]{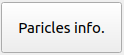} button. Particle size information appears in a small dialog box, as shown in Figure~\ref{ref:post_processing_1}. For the composition profile analysis, users can choose a point by clicking on the microstructure in the main \mybox{$\mu$2mech} window. For example, the coordinate of the chosen point is (X, Y)=(166,359) in Figure~\ref{ref:post_processing_1}. The composition profile is drawn along a horizontal or vertical line passing through the selected point. As shown in Figure~\ref{ref:post_processing_1}, this example draws the composition profile along the horizontal (x) direction. Clicking on the \includegraphics[width=0.15\textwidth]{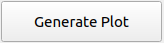} button creates the composition plot along the designated direction, as illustrated in Figure \ref{ref:post_processing_1}. Users can visualize the time evolution of the composition profile by clicking the left/right arrow key or sliding the slider (located at the bottom of the \textit{Post Processing} subwindow). Users can also save the composition profile(s) using the ``Save Plots'' or ``Save all plots'' options and make an animation using the ``Export Animation'' option by clicking the respective \includegraphics[width=0.08\textwidth]{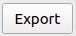} buttons in the \mybox{Post Processing} dialog box [Figure \ref{ref:post_processing_1}].

The property prediction feature of the post-processing module of $\mu$2mech allows users to predict elastic properties by performing Finite Element Analysis (FEA) directly over the simulated microstructure (obtained via phase field modeling). The current version of $\mu$2mech allows the prediction of elastic properties from  $2$D microstructures containing only two phases, which is accomplished by integrating $\mu$2mech with OOF$2$, an open-source object-oriented finite element method software developed by NIST, USA\cite{OOF2}.

First, let us describe the general operation of OOF$2$ briefly. OOF$2$ takes a microstructure and allows users to identify phases based on the constituent phases' pixel values (or gray-scale values). Next, OOF$2$ needs the elastic properties of the individual phases to be added by the user. If the phases are isotropic, only two material properties, namely elastic modulus and Poisson's ratio, are needed as input, while based on the degree of anisotropy, multiple components of the elastic stiffness matrix, known as the elastic constants, are needed. Once the phases are identified, it allows the users to build a ``skeleton" that identifies the boundaries of the phases so that a mesh with elements and nodes conforming to the phase boundaries can be generated. Building the skeleton requires initial input from the user regarding the number of elements along the horizontal edges (or x-direction) and along the vertical edges (or y-direction). Mesh generation involves further refinement of the elements so that the elements conform to the phase boundaries, as described before. Before meshing, OOF$2$ allows users to choose triangular or quadrilateral elements, similar to those found in ABAQUS, a commercial finite element software. Mesh generation is followed by the assignment of boundary conditions. Displacement boundary conditions are imposed on the microstructure's left, right, top, and bottom edges. The displacements are applied in terms of pixels. This is followed by solving the finite element model using the default iterative solver available with OOF$2$. Post-processing involves a calculation of the area average of the required stress and strain components, from which the elastic properties can be calculated. 

These operations are automated via a python code that is integrated within the $\mu$2mech which calls OOF$2$ (which should be separately installed), loads the specified virtual phase-field model-generated microstructure, applies the boundary conditions for a simulated tensile test (along the horizontal axis or the vertical axis), solves the finite element model, and performs post-processing operations, calculating the elastic properties and saving these properties in a file at a chosen destination. User inputs are required for material properties of the constituent phases of the microstructure, for the selection of the type of element, for quantifying the displacements applied on the boundaries of the microstructure, and for the choice of destination for saving the output files. It is important to note here that only a small set of features of OOF$2$ software are being utilized by $\mu$2mech through a Python code that automates virtual tensile tests in an elastic regime for a microstructure containing two phases, one of which is continuous, while the other is discontinuous. To explore the full features of OOF$2$, it is recommended that the users export the generated microstructural image files and open these files directly in OOF$2$, i.e., outside the environment of $\mu$2mech. The following steps must be followed to initiate the evaluation of the elastic properties (namely elastic moduli and Poisson's ratio) in $\mu$2mech.

\begin{figure}
    \centering
    \includegraphics[width=0.5\textwidth]{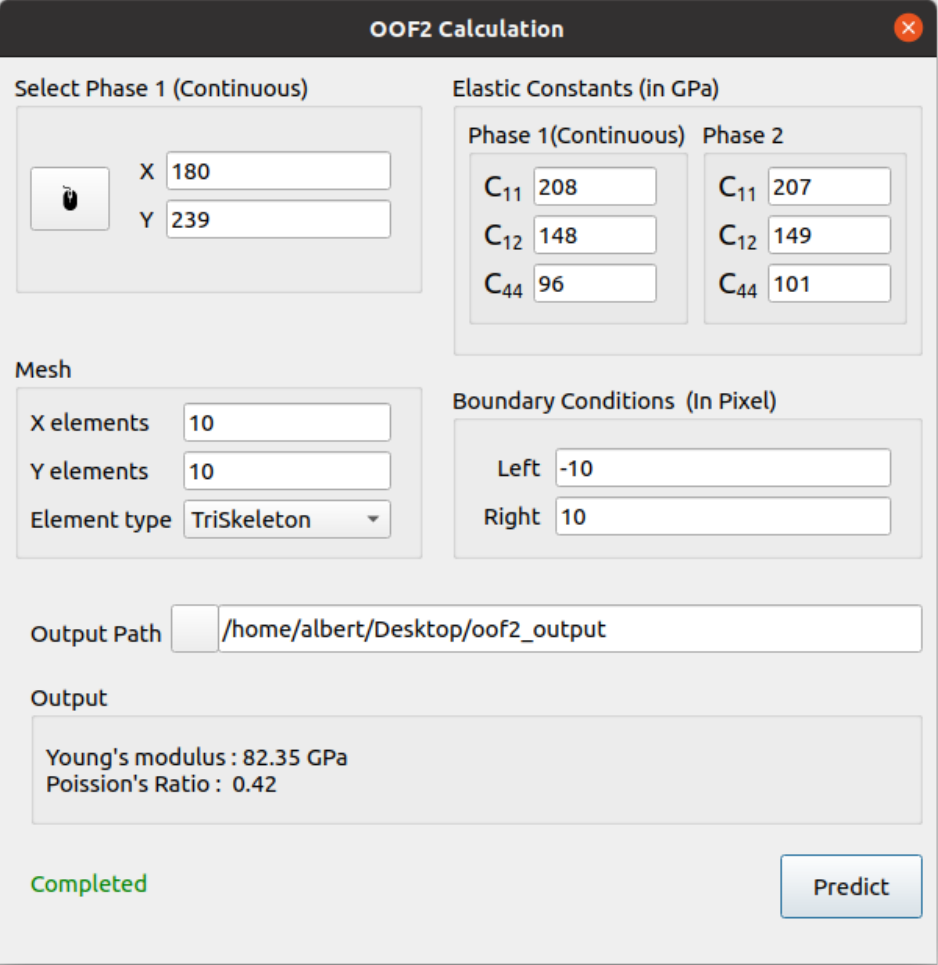}
    \caption{Dialog box related to OOF2 Calculation.}
    \label{ref:post_processing_2}
\end{figure}

First, the users click on the \includegraphics[width=0.1\textwidth]{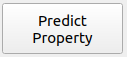} button in the \mybox{Post Processing} dialog box, which opens another dialog box \mybox{OOF$2$ Calculation}, as shown in Figure \ref{ref:post_processing_2}. In this dialog box, the users click on the mouse-shaped button \includegraphics[width=0.04\textwidth]{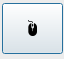}, which allows the users to click on the continuous phase in the main \mybox{$\mu$2mech} window displaying the microstructure. $\mu$2mech uses the \textit{burn} and \textit{invert} options associated with pixel-selection methods of OOF$2$ for phase identification. The \textit{burn} operation works by selecting contiguous pixels of the same color (i.e., RGB or gray-scale pixel values) for one phase until it encounters pixels with different colors (or pixel values). The \textit{invert} option can be used only if the second phase is discontinuous, and it selects pixels other than those stored using the \textit{burn} operation of the continuous phase. These operations happen in the background. On clicking on the continuous phase, only the $x$ and $y$ coordinate values of the point of mouse-click are displayed in the \mybox{OOF$2$ Calculation} dialog box next to the \includegraphics[width=0.04\textwidth]{Inline_Images/mouse_shaped_btn.png} button. 
	
Next, the users enter the elastic constants for both phases under the header ``Elastic Constants (in GPa)''. At present, $\mu$2mech can only work with microstructures where one or both phases are either isotropic or cubic in terms of material symmetry. For an isotropic phase, the values of the elastic constants can be calculated from Young's modulus and Poisson's ratio. Elastic constant values are either obtained from the experimental data or from \textit{ab initio} calculations. Determination of elastic constants from \textit{ab initio} calculations have been discussed briefly in Appendix~\ref{secelastic}.
		
\begin{figure}[t]
    \centering
    \includegraphics[width=1.0\linewidth]{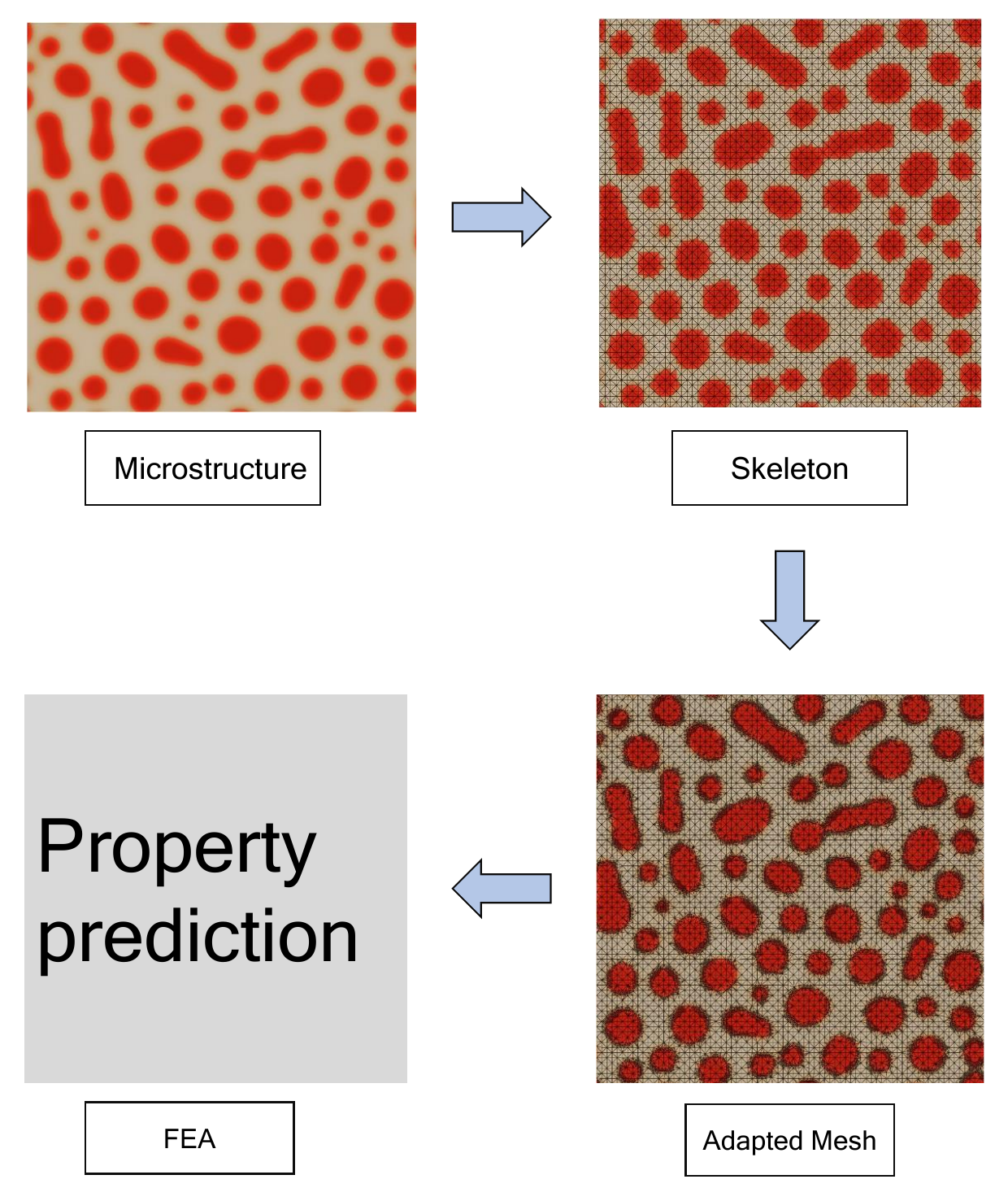}
    \caption{Mesh adaption in OOF2 from input skeleton through skeleton modification tools. }
    \label{ref:adapted_mesh}
\end{figure}

Next, the users enter the number of elements along the horizontal and vertical edges of the microstructure in the boxes beside ``X elements'' and ``Y elements'' under the header ``Mesh''. This is followed by choosing the element type by clicking on the drop-down menu for ``Element type$\downarrow$'' under the header ``Mesh'' and selecting ``TriSkeleton'' or ``QuadSkeleton''. OOF$2$ allows only two geometries of elements, viz. triangular and quadrilateral, with triangular elements (``TriSkeleton'') as the default option. This initiates the process of building the skeleton and is followed by meshing. OOF$2$ generates an adapted mesh using its skeleton modification tools to account for the microstructure's geometrical heterogeneity. An example of meshing has been illustrated in Figure~\ref{ref:adapted_mesh}.

Next, the users enter the boundary conditions for a simulated tensile test (in elastic regime) along the horizontal (or $x$) direction under the header ``Boundary Conditions (in Pixel)''. Displacements are applied in terms of number of pixels in OOF$2$. OOF$2$ considers displacements along the $+x$-direction positive and displacements along the $-x$-direction negative. Hence, for a tensile test along the horizontal direction, normal displacements need to be applied along the outer normals, i.e., perpendicular to the left and right edges of the microstructure. Hence, the normal displacement applied on the left edge has a negative value, while that on the right edge has a positive value. It is recommended that the displacements along the left and right edges should have the same magnitude (i.e., absolute value without the positive or negative signs). The users should enter a negative value for the displacement in the box beside ``Left'' and a positive value (of the same magnitude) beside ``Right''. 

The users must specify the output path to store the output files from OOF$2$. OOF$2$ generates files with area-averaged stress and strain component values, from which the elastic properties are calculated. Finally, clicking on the \includegraphics[width=0.07\textwidth]{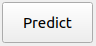} button initiates the FEM calculations. At this point, OOF$2$ starts running in the background. While the calculation is running, the status bar at the bottom left will show the status ``Performing calculation", and upon completion, output files for stress, strain, and mesh are generated at the specified output path. Once OOF$2$ operations are completed, the computed elastic properties, i.e., Young's modulus and Poisson's ratio, are displayed at the bottom of the dialog box.

\section{Impact}
Replacing the age-old trial-and-error method for developing new alloys using the ICME framework requires the integration of multiple time and length-scale simulations. The $\mu2$mech package integrates microstructure prediction via the mesoscale phase field method to the microscale finite element analysis for mechanical property prediction. The package can quantitatively predict by taking inputs from CALPHAD and \textit{ab initio} calculations. Such a multiscale package will be handy for the materials science and engineering community, especially those working in alloy development and microstructure engineering for modulating mechanical properties.

Some salient features of the $\mu2$mech package are the following. The software's architecture, including open-source libraries, makes it easy to use and accessible to a broader range of users. The graphical user interface makes it very easy to run the code, even for beginners. At the same time, the codebase is written in such a way that newer components can easily be added. Its modular approach allows flexibility to incorporate different phase field models, making it a valuable tool for advanced users as well. With the increasing application of artificial intelligence (AI) in phase-field modeling\cite{owais2023}, a package like $\mu2$mech will be convenient for relatively effortless microstructure generation in bulk, which can be further used for training machine learning (ML) models.

\section*{Acknowledgements}
RM and SB are thankful for financial support received from C-DAC Project No. Meity/R\&D/HPC/2(1)/2014. Authors acknowledge National Super Computing Mission (NSM) for providing computing resources of ``PARAM Sanganak'' at IIT Kanpur, which is implemented by CDAC and supported by the Ministry of Electronics and Information Technology (MeitY) and Department of Science and Technology (DST), Government of India. We also acknowledge the HPC facility provided by CC, IIT Kanpur and ICME National Hub, IIT Kanpur.

\appendix
\section{Installation Instructions}
\label{secinstall}
Follow the steps below to install and run \texttt{mu2mech} on your linux system:

\begin{enumerate}
  \item Download and extract \texttt{mu2mech} \\
  \textit{wget https://github.com/mu2mech/mu2mech} \\
  \textit{unzip mu2mech.zip} \\
  
  \item Create and activate a Python environment:
  \textit{python3 -m venv mu2mech-env} \\
  \textit{source mu2mech-env/bin/activate} \\
  
  \item Install packages: \\
  \textit{pip install -r requirements.txt} \\
  \textit{sudo apt install ffmpeg} \\
  
  \item Run the program: \\
  \textit{python3 mu2mech.py}
\end{enumerate}

Troubleshooting: Following are some common errors one may encounter during the installation or execution of \texttt{mu2mech} and suggested remedies.\\

\begin{itemize}
  \item \textbf{ImportError: libOpenGL.so.0: cannot open shared object file: No such file or directory} \\
  Resolve this issue by installing the \texttt{libopengl0} package: \\
  \textit{sudo apt install libopengl0 -y} \\
  
  \item \textbf{qt.qpa.plugin: Could not load the Qt platform plugin "xcb" in "" even though it was found.} \\
  Reinstall the \texttt{libxcb-xinerama0} package to fix this problem: \\
  \textit{sudo apt-get install --reinstall libxcb-xinerama0}
\end{itemize}

\section{CALPHAD integration}
\label{seccalphad}
\begin{table*}
\centering 
\caption{Table showing the ThermoCalc-extracted dataset essential for estimating the Gibbs energy plot as per Equation (2).}
\begin{tabular}{|c|ccccc|cc|}
\hline
Temp. (\degree C) & \multicolumn{5}{c|}{Coefficients} & \multicolumn{2}{c|}{Binodal points} \\
\cline{2-8}
& $A_1$ & $A_2$ & $A_3$ & $A_4$ & $A_5$ & point 1 & point 2 \\
\hline
200 & 1 & -2.046 & 1.0964 & -0.051 & 0.0006 & 0.025 & 0.998 \\
220 & 1 & -2.046 & 1.0964 & -0.051 & 0.0006 & 0.025 & 0.998 \\
240 & 1 & -2.046 & 1.0964 & -0.051 & 0.0006 & 0.025 & 0.998 \\
260 & 1 & -2.041 & 1.1339 & -0.0943 & 0.0021 & 0.0475 & 0.973 \\
280 & 1 & -2.041 & 1.1339 & -0.0943 & 0.0021 & 0.0475 & 0.973 \\
300 & 1 & -2.0314 & 1.1339 & -0.0943 & 0.0021 & 0.0475 & 0.973 \\
320 & 1 & -2.0314 & 1.1339 & -0.0943 & 0.0021 & 0.0475 & 0.973 \\
340 & 1 & -2.041 & 1.1789 & -0.1403 & 0.0047 & 0.0725 & 0.948 \\
360 & 1 & -2.041 & 1.1789 & -0.1403 & 0.0047 & 0.0725 & 0.948 \\
380 & 1 & -2.091 & 1.2779 & -0.1933 & 0.0085 & 0.0975 & 0.948 \\
400 & 1 & -2.041 & 1.2214 & -0.1837 & 0.0081 & 0.0975 & 0.923 \\
420 & 1 & -2.092 & 1.3212 & -0.2375 & 0.0129 & 0.123 & 0.923 \\
440 & 1 & -2.092 & 1.2633 & -0.2255 & 0.0122 & 0.123 & 0.898 \\
460 & 1 & -2.042 & 1.3599 & -0.278 & 0.0177 & 0.148 & 0.898 \\
480 & 1 & -2.092 & 1.3962 & -0.316 & 0.0228 & 0.173 & 0.873 \\
500 & 1 & -2.092 & 1.4299 & -0.3513 & 0.0282 & 0.198 & 0.848 \\
\hline
\end{tabular}
\label{ref:tc_table}
\end{table*}
The system's free energy strongly influences the equilibrium compositions of the phases and subsequent microstructure evolution. CALPHAD (Calculation of Phase Diagrams) tools, such as Thermo-Calc, OpenCalphad, and Pandat, have become invaluable for obtaining thermodynamic data for studying alloy systems. These tools employ computational methods to calculate phase equilibria, phase diagrams, and thermodynamic properties based on a combination of experimental data and theoretical models.

One extracts the data points from the free energy curve and constructs a common tangent between individual valleys to determine the equilibrium compositions. The equilibrium points represent the compositions at which the system is in a minimum free energy state. These points are crucial for characterizing the stable phases and predicting phase transformations in the alloy system. One can identify the equilibrium points using mathematical tools such as OCTAVE or Thermo-Calc single-point calculation. One can approximate the free energy by a simple 4$^{th}$-order polynomial equation:
\begin{equation}
G=A(c-c_1)^2(c-c_2)^2
\end{equation}
where $c_1$ and $c_2$ are the equilibrium compositions and $A$ is a constant that influences the barrier height of the free energy curve. This polynomial equation provides a mathematical representation of the free energy curve and enables further analysis and calculations. One can express the polynomial equation as:
\begin{equation}
\label{eq2}
G=A_1c^4+A_2c^3+A_3c^2+A_4c+A_5
\end{equation}
where $A_1=A$, $A_2=A(c_1+c_2)$, $A_3=A(c_1^2+c_2^2+4c_1c_2)$, $A_4=-2Ac_1c_2(c_1+c_2)$, and $A_5=Ac_1^2c_2^2$. These polynomial coefficients allow for the direct calculation of the free energy values at different compositions within the alloy system.

The scaled free energy is utilized in our simulation, where the free energy equation above is divided by $A$. This scaling is necessary to ensure that the magnitude of the free energy values is appropriate for numerical calculations and simulations. For example, at a specific temperature, the equilibrium compositions $c_1$ and $c_2$ are determined based on the analysis of the free energy curve. The polynomial coefficients $A_1$, $A_2$, $A_3$, $A_4$, and $A_5$ can then be calculated using the Equation~\ref{eq2}. This approach can be extended to other temperatures of interest, allowing for the construction of polynomial equations that describe the free energy behavior at those temperatures. Integrating these polynomial equations with a phase field model makes it possible to simulate and analyze the alloy system's microstructure evolution and phase transformations under different thermodynamic conditions.

\section{Determination of Elastic Constants}
\label{secelastic}
The assessment of elastic constants begins with the initial estimation of the bulk modulus, achieved through the utilization of the Birch-Murnaghan equation of state,
\begin{equation}
    E(V) = E(V_0) + \frac{B}{B^{\prime}} \left[ \frac{(V/V_0)^{B^{\prime}}}{B^{\prime}-1} + 1 \right] - \frac{B V_0}{B^{\prime}-1}.
\end{equation}
Energy as a function of volume is calculated using Density Functional Theory (DFT).

To determine the values of $C_{11}$ and $C_{12}$, the unit cell is subjected to deformation. In the case of cubic materials, the following volume-preserving strain is applied on the conventional unit cell,
\begin{equation}
    \varepsilon =
    \begin{pmatrix}
        \delta & 0 & 0 \\
        0 & -\delta & 0 \\
        0 & 0 & \frac{\delta^2}{1-\delta^2}
    \end{pmatrix}.
\end{equation}
This transformation changes the cubic cell to an orthorhombic one.
The energy of the deformed cell satisfies the following equation,
\begin{equation}
    E(\delta) = E(0) + (C_{11} - C_{12})V\delta^2.
\end{equation}
$E(0)$ is the energy of the pristine structure, and $E(\delta)$ is the energy of the deformed structure, obtained from DFT calculations. Subsequently, by recognizing that $B = \frac{C_{11} + 2C_{12}}{3}$, the values of $C_{11}$ and $C_{12}$ can be derived.

The evaluation of $C_{44}$ involves the utilization of the following strain matrix,
\begin{equation}
    \varepsilon =
    \begin{pmatrix}
        0 & \delta/2 & 0 \\
        \delta/2 & 0 & 0 \\
        0 & 0 & \frac{\delta^2}{4-\delta^2}
    \end{pmatrix}.
\end{equation}
Using the following equation, one can obtain the value of $C_{44}$,
\begin{equation}
    E(\delta) = E(0) + \frac{1}{2}C_{44}V\delta^2,
\end{equation}
where $E(\delta)$ is the energy of the deformed structure,  obtained from DFT calculations.
\bibliographystyle{elsarticle-num}  
\bibliography{references}
\end{document}